%% file: ex_article.tex
\documentclass[hidelinks,onefignum,onetabnum]{siamart250211}

\input{ex_shared}

\ifpdf
\hypersetup{
  pdftitle={Temperature induced tipping in a two-box ocean circulation model},
  pdfauthor={J. Noory}
}
\fi

\usepackage{amsfonts, amsmath,amssymb, comment}

\begin{document}

\maketitle

\begin{abstract}
 Climate tipping points are critical thresholds in Earth's climate system where a small change can cause abrupt and potentially irreversible shifts towards a new state. Tipping points in the Atlantic Meridional Overturning Circulation (AMOC) are of much scientific concern because of the large-scale impacts on Earth's climate. A two-box model representing the AMOC, introduced by oceanographer Paola Cessi (1994), revealed a multistable system, and was used to develop a tipping analysis framework to capture critical thresholds under freshwater forcing. Sustained increases in planetary energy uptake motivate the inspection of the model, shifting the focus to thermal dynamics. In this paper, we demonstrate that tipping can occur in the model where the temperature gradient acts as the forcing parameter instead of the freshwater forcing parameter. 
\end{abstract}




\section{Introduction}

   The Intergovernmental Panel on Climate Change (IPCC) has defined a tipping point as “a critical threshold beyond which a system reorganizes, often abruptly and/or irreversibly” \cite{IPCC}. Critical thresholds are system characteristics that delimit where the qualitative shifts of state occur due to small changes in forcings. Dynamical systems tools are often applied to study where these critical thresholds occur, and they can point to phenomena like hysteresis, where returning to the initial forcing does not necessarily reverse the tipping. 

    The Atlantic Meridional Overturning Circulation (AMOC) is a part of the global ocean current, a mechanism by which the ocean circulates heat and regulates regional climate patterns. The AMOC functions by sending warm, salty water from the tropics northward toward the Arctic. As this water reaches colder regions, it cools, becoming denser, and sinks into the deep ocean. This dense water then flows southward at depth, eventually warming again as it moves toward the equator. As it warms, its density decreases, and it rises back to the surface, completing a large-scale circulation loop. This process is slow, taking roughly 1,000 years for a parcel of water to complete the full cycle. However, global warming disrupts this system: higher temperatures reduce the cooling and densification of water in polar regions, weakening the sinking process and slowing down the overall circulation.

    The observed slowing of the circulation rate \cite{Caesar2018} has raised concern for a potential tipping event, making the AMOC a focus of many scientific studies, approached observationally, computationally, and conceptually.. The AMOC has been linked to abrupt climate transitions in the paleoclimate record. For example, fluctuations in the AMOC drive the rapid warming phases of Dansgaard-Oeschger events by altering heat transport to the North Atlantic, triggering abrupt climate shifts. To study tipping events, researchers often generate AMOC simulations on highly complex Global Climate Models (GCMs). For example, Ditlevsen and Ditlevsen \cite{Ditlevsen2023} provided a data-driven estimator for time of tipping using statistically significant early warning signals, like increases in autocorrelation and variance, also known as classical early warning signals. Another approach by van Westen \cite{vanWesten2024} was to develop a physics-based early warning signal using numerically estimated minimums of the overturning component of freshwater transport due to the AMOC. Whereas these studies focus on freshwater forcing impacting the density through salinity changes, other studies focus on density changes due to temperature-induced tipping in a double well potential, often in the framework of noise-induced tipping \cite{Chapman2024}. Although rigorously approached, the observational and computational studies are often based on datasets with preprocessing and missing data filling methods which introduce a large amount of uncertainties, too large to predict a time to tip \cite{Ben-Yami2023}.

\begin{figure}[h]
        \centering
        \vspace{-1em}
        \includegraphics[scale=.4]{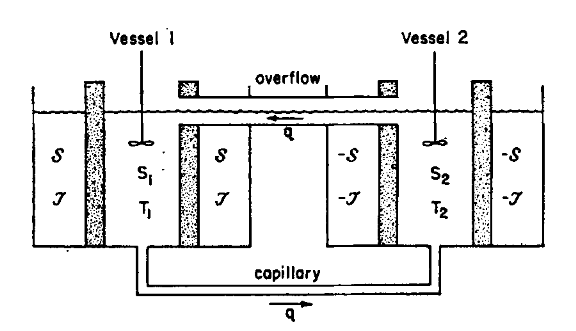}
        \caption{Schematic of the Stommel's model.   ; Figure 5 of \cite{Stommel1961}.}
        \label{fig: Stommel}
        \vspace{-1em}
    \end{figure}

    An alternate lens to study tipping is through conceptual circulation models, often introducing minimally-complex box models. The first ocean circulation conceptual model was introduced by oceanographer Henry Stommel in his 1961 paper \cite{Stommel1961} using a two-box model representing a southern equatorial ocean and a northern ocean. The dynamics of this box model illustrated how temperature and salinity combine to produce the thermohaline circulation. The dynamic variables are the temperature and salinity differences between the boxes. The density in each box is a function of temperature and salinity, and the density difference between the two boxes drives the flow between the boxes, shown by variable $q$ in figure \ref{fig: Stommel}.  Stommel showed that the system has two stable nodes, one where the flow rate $q$ is dominated by temperature, and the other by salinity. 

\begin{figure}[H]
        \centering
        \vspace{-1em}
        \includegraphics[scale=.4]{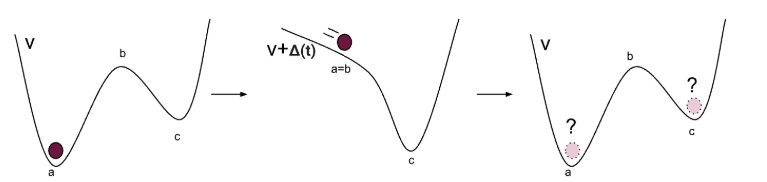}
        \caption{Depiction of tipping framework using the potential landscapes.}
        \label{fig: tipping framework}
        \vspace{-.25em}
    \end{figure}

    In 1994, Cessi \cite{Cessi1994} simplified Stommel's model with the goal of more easily analyzing the dependence on parameters.  By hypothesizing that temperature dynamics happen much faster than salinity dynamics, she reduced Stommel's box model from a system of two ordinary differential equations to a single ordinary differential equation of the salinity dynamics, reinterpreting the dynamics as a potential landscape. The double well potential of the salinity dynamics demonstrate the two alternate, stable modes of Stommel's model, one corresponding to a weaker circulation dominated by temperature and the other a stronger one dominated by salinity, corresponding to the local minima of the potential, as illustrated in figure \ref{fig: tipping framework}.

    Cessi was able to reduce Stommel's two variables down to one by assuming that the temperature difference between the two boxes equilibrated much faster than the salinity difference. Taking the temperature difference as fixed at its equilibrium value, she was able to focus on the salinity difference to see how its dynamics changed as the freshwater influx was changed. Given the rapidly changing global temperatures and the polar amplification effect, it makes sense to explore Cessi's model using the equilibrium temperature difference as the parameter of interest.
    
\vspace{1cm}
\section{Background}
    We start with a detailed description of Cessi's model before describing the effect of focusing on the equilibrium temperature difference as the parameter of interest.
    
    \subsection*{The Model}
    Assume there are two well-mixed liquid reservoirs with the following external forcing: a freshwater influx into the compartments as well as an ambient temperature relaxation target, which may differ for each box. Assume that density is described by a linear combination of temperature and salinity with characteristic contraction and expansion rates. Finally, assume that both reservoirs interact through an exchange that is driven by the density difference between boxes. This is Stommel's box model \ref{Stommel Model}, and it captures thermohaline circulation. A particularly clear explanation of this model is given in Kaper and Engler's textbook \cite{HansKaper}, which helped guide the exposition presented here.
    
     \begin{align}
        \frac{\Delta T}{dt} &= -t_r^{-1}(\Delta T - \theta ) - Q(\Delta\rho) \Delta T  \notag \\ 
        \frac{\Delta S}{dt} &= \frac{F(t)}{H}S_0 - Q(\Delta\rho) \Delta S  \\ \label{Stommel Model}
        \Delta \rho &= \alpha_S \Delta S - \alpha_T \Delta T \notag
    \end{align}

    The model expresses thermohaline circulation between two compartments representing a two-box ocean for northern and southern latitudes with a system of ordinary differential equations. The system tracks the time evolution of temperature differences $\Delta T$ and salinity difference $\Delta S$ between the boxes. Temperature and salinity differences are forced by different processes. The temperature forcing process in this model is Newton's law of cooling, represented by a relaxation of $\Delta T$ towards an ambient temperature difference equilibrium, $\theta$. The rate at which the boxes equilibrate to $\theta$ is $t_r^{-1}$, the characteristic relaxation timescale. The salinity difference is forced with a prescribed freshwater flux $F(t)$, representing the imbalances between evaporation and precipitation plus freshwater runoff from rivers and glaciers into the North Atlantic. $F(t)$ acts indiscriminately on the whole box of depth $H$ with reference salinity state $S_0$.

    The interaction between the boxes is described by exchange function $Q$, which is driven by the density difference between the two, $\Delta \rho$. The density difference, $\Delta\rho$, depends on a linear combination of temperature and salinity differences through thermal expansion coefficient $\alpha_T$ and haline contraction coefficient $\alpha_S$. The exchange function $Q$ is of the form
    \begin{align}
        Q(\Delta \rho) = t_d^{-1} + M^{-1} q(\Delta \rho)^2, \label{eqn: exchange function}
    \end{align}

    where $M$ is the volume of the box and $t_d$ is the characteristic timescale of salinity diffusion. The exchange in equation \ref{eqn: exchange function} differs from Stommel's exchange function, and Cessi justifies the choice by referencing previous work \cite{CessiYoung} which justifies that the form of convection in equation \ref{eqn: exchange function} leads to nonlinear exchange, as in Stommel's model. Mathematically, Cessi's exchange form ensures that exchange between boxes occurs smoothly, and Cessi notes that it is crucial to the expression of three equilibria in the model, two of which are stable.

    To facilitate the analysis, the following change of variables is used to nondimensionalize the system. The new variable $x$ tracks a dimensionless temperature difference, and the variable $y$ tracks the dimensionless salinity difference. Time $t$ represents a scaled time expanded to the characteristic diffusion timescale.
    \[ x = \frac{\Delta T}{\theta}, \ \ \ y = \frac{\alpha_S \Delta S}{\alpha_T \theta}, \ \ \ t = t_d t'.\]
    
    Then equations \ref{Stommel Model} become
    \begin{align}
        x' &= -\alpha(x-1) - x(1+\mu^2(x-y)^2) \label{nondimensional Cessi model}\\
        y' &= p(t) -y(1+\mu^2(x-y)^2). \notag
    \end{align}
    
   Here $\alpha=\frac{t_d}{t_r}$, describes the ratio of characteristic timescales of diffusion to relaxation. Cessi introduced estimates $t_d \approx 219$ years and $t_r \approx 25$ days.  The parameter $\mu^2 = qt_d(\alpha_T \theta)^2/M$ describes the ratio of diffusive timescale to advective timescale. In this parameter expression, the transport of total volume $M$ is computed using the typical width of the Western boundary current rather than the entire width of the North Atlantic so to keep consistent with the assumption that present day transport is dominated by the advective process. Lastly, $p$ is the nondimensional freshwater flux with $p(t) = \frac{\alpha_S S_0 t_d}{\alpha_T \theta H} F(t)$.  The freshwater flux is approximated by $p(t) = \bar{p} + p'(t)$, where $\bar{p}$ is a time averaged forcing flux and $p'(t)$ is not a derivative in time, but a white noise stochastic component representing a pulse of freshwater. In the results below, we will not use $p'(t)$, but instead will treat the freshwater forcing only as a fixed time-averaged forcing. With an average estimate for the North Atlantic fresh water flux $\bar F = 2.3$ m yr$^{-1}$ and $\theta = 20^{\circ}$C, then $p\approx 1$. The following table demonstrates the parameter choices in the model.

\vspace{0.5cm}   
    \begin{center}
    \begin{tabular}{ |c|c|c| } 
     \hline
     Parameter & Dimensionless Value & Representation \\ 
     \hline
     $\alpha$ & 3600 & ratio of diffusive to relaxation timescales \\ 
     $\mu^2$ & 6.2 & ratio of advective to diffusive timescales \\ 
     $p$ & 1 & nondimensional freshwater flux \\ 
     \hline
    \end{tabular}
    \end{center}
\vspace{0.5cm}    

    Cessi further reduces the two dimensional model by noticing that the very large parameter $\alpha$ induces a distinction in the timescales of temperature relaxation and salinity diffusion. The fast dynamics can be approximated by holding them constant at their critical value over time, which effectively reduces the model down to one governing equation. Therefore Cessi contributes a Stommel's model simplification, where equation \ref{one dimensional cessi model} is the one dimensional ordinary differential equation of the describing the circulation.
    \begin{align}
        x&\approx 1  \notag \\
        y' &= -y(1+\mu^2(y-1)^2) -p(t) + \mathcal{O}(\epsilon^2) \label{one dimensional cessi model}
    \end{align}
    
     Cessi pointed out that the governing dynamics of the salinity difference can be understood using potential functions by reinterpreting $\frac{dy}{dt'} = -V'(y)$, where $V(y)$ is the derivative of a Lyapunov function or a potential function in a neighborhood of the stable equilibria.  The critical points of the potential function are indeed the equilibrium points of $\frac{dy}{dt'}$, where the minima correspond to the stable equilibria and the maxmia correspond to the unstable equilibria. One can integrate the one dimensional model \ref{one dimensional cessi model} with respect to $y$ to get the potential function $V(y)$
    \begin{align} 
    V(y) = \mu^2 (\frac{y^4}{4}  - \frac{2y^3}{3} + \frac{y^2}{2}) + \frac{y^2}{2} - py  \label{eqn: Cessi potential}
    \end{align}

\begin{figure}[H]
        \centering
        \includegraphics[width=0.8\linewidth]{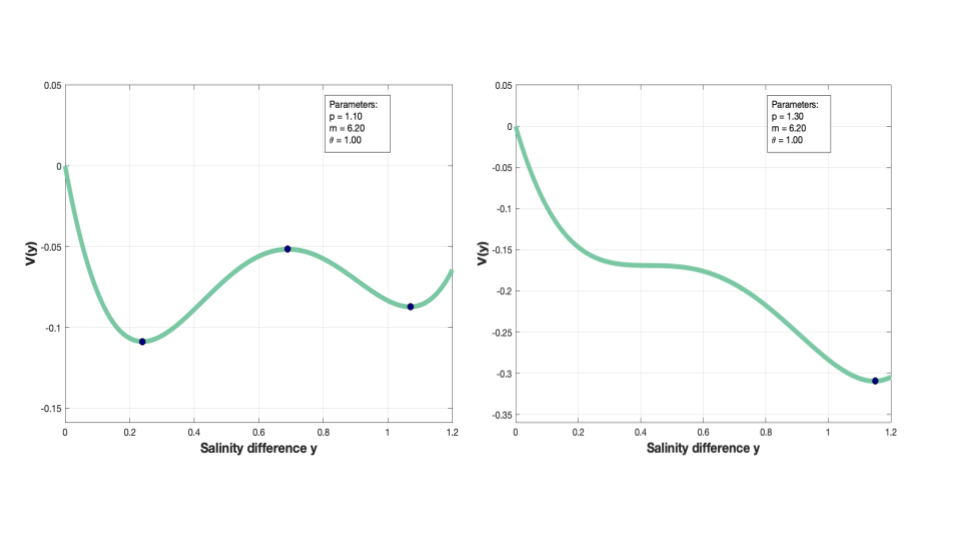}
        \caption{Depiction of potential landscapes as a function of the salinity difference $y$ under two parameter regimes. The panel on the left expresses the potential under freshwater forcing $p=1.1$, exchange rate 6.2, and temperature equilibrium $\theta=1$. Two local minima correspond to stable salinity difference states at 0.24 and 1.07, and one local maximum, corresponds to unstable state at 0.69. The panel on the right expresses the potential under increased freshwater forcing, $p=1.3$, with other parameters held the same as left panel. The increase in freshwater results in a structural change to only one local minima corresponding to a stable salinity difference state greater than 1.07.  }
        \label{fig:Cessi-potentials}
    \end{figure}

    It is easy to visualize the system’s stability and inspect how it changes in response to freshwater forcing using potential functions. Forcing the system at level $p(t)\equiv 1.3$, the critical forcing is reached and the potential landscape changes from a multistable landscape (see left panel of figure \ref{fig:Cessi-potentials}), in the following way. The left most stable state, corresponding to a stronger circulation rate driven by salinity, collides with the unstable state (local maximum), annihilating both equilibria and inducing a globally stable landscape. A critical threshold has been reached in the freshwater forcing parameter $p$. In this parameter regime, flows will be influenced towards the alternate stable state, corresponding to the stronger circulation rate. 

     This framework illuminates critical changes to the dynamic  structure of density driven circulation subject to freshwater forcings, occurring at a timescale that is in the order of centuries. However, rapid increases in shallow ocean heat content and global mean temperatures justify the inspection of AMOC tipping subject to regime changes in temperature. In what follows, we analyze emergent dynamic bifurcations subject to temperature forcing on Cessi's ocean circulation model.

\vspace{1cm}
\section{Main results}
\label{sec:main}

We begin by demonstrating that the ocean circulation model \ref{Stommel Model} can be reduced to a one-dimensional salinity equation with an explicit temperature parameter: the equilibrium temperature gradient $\theta$. Through a scaling and dimensional reduction, we obtain a dimensionless salinity variable while retaining $\theta$ as a physically meaningful parameter. Performing the following change of variables on equations \ref{Stommel Model},

\[x \equiv \Delta T, \ \ y \equiv \frac{\alpha_S \Delta S}{\alpha_T}, \ \  t = t_d t'\]

the model transforms to 

\begin{align}
Q(\Delta \rho) &=\frac{1}{t_d} + \frac{q\alpha_T^2}{V}(y-x)^2 \\
\frac{dx}{dt'} &=\frac{-t_d}{t_\alpha} (x - \theta ) - x [1 + \frac{t_d q\alpha_T^2}{V}(y-x)^2]\\
\frac{dy}{dt'} &= \frac{\alpha_S t_d}{\alpha_T} \frac{F(t)}{H}S_0 -y[ 1 + \frac{t_d q\alpha_T^2}{V}(y-x)^2]
\end{align}

We intentionally preserve temperature units (Celsius) in the model so that variations in $\theta$ retain its physical interpretation, the temperature gradient between low latitude and high latitude oceans. Thus, $x = \Delta T$ retains units of Celsius, and time is nondimensionalized using the same scaling as Cessi's analysis $t=t_dt'$.

One can notice that the parameter $\alpha = \frac{t_d}{t_r}$ is very large. Let $\epsilon = \frac{1}{\alpha}$, then we can rewrite the above system as 

\begin{align}
\epsilon \frac{dx}{dt'} &= (x - \theta ) - \epsilon x [1 + m^2(y-x)^2]\\
\epsilon \frac{dy}{dt'} &= \epsilon p - \epsilon y[ 1 + m^2(y-x)^2]
\end{align}

where $m^2=\frac{t_d q\alpha_T^2}{V}$ and $p'(t) = \frac{\alpha_S t_d}{\alpha_T} \frac{F(t)}{H}S_0$ and the parameter $\epsilon <<1$. 

We can clearly identify the system as a fast-slow system by taking the limit as $\epsilon$ approaches 0. The nontrivial relation that holds is $x = \theta$, which is the critical manifold of the fast variable $x$. Because the temperature difference $x$ evolves at a fast timescale relative to salinity difference $y$, we can further simplify the analysis of the system by projecting the salinity dynamics onto the critical manifold, where the fast variables are assumed to be at quasi-equilibrium $x = \theta$. The one dimensional system is as follows.

\begin{align}
    x &\approx  \theta\\
\frac{dy}{dt} &= p(t) -y[ 1 + m^2(y-\theta)^2] \label{autonomous salinity difference dynamics}
\end{align}

In this view, the model describes the salinity dynamics in the timescale of temperature (Celsius), with three parameters. The parameter $p(t)= \frac{\alpha_S t_d}{\alpha_T} \frac{F(t)}{H}S_0$ describes a freshwater forcing. The parameter $m^2 =\frac{t_d q\alpha_T^2}{V}$ corresponds to the rate of exchange between boxes with respect to the ratio of the diffusive timescale $t_d$ to the advective one. The salinity dynamics are described with an explicit temperature parameter $\theta$, which is the forcing parameter of interest: the relaxation target of the fast variable $\Delta T$. It represents the ambient temperature gradient between the high and low latitudes surrounding each ocean box. 

The question is, "Are there any dynamic changes to system structure as the temperature gradient $\theta$ varies?" As a control system, we consider the same autonomous system in Cessi's original analysis \cite{Cessi1994} that provides the control double well structure. Fix the freshwater forcing to maintain the double well structure with $p(t) \equiv p^* = 1.1$ and the exchange rate at $m^2 =6.2$. Notice that $\theta=1$ is the parameter regime in the control system.

In what follows, we demonstrate a saddle-node bifurcation as the temperature gradient parameter $\theta$ varies using Cessi's potential function framework.

\subsubsection*{The Potential Function}
One might notice that the governing dynamics of the salinity difference can be understood as $\frac{dy}{dt} = -V'(y)$, where $V(y)$ is the derivative of a Lyapunov function or a potential function in a neighborhood of the stable equilibria.  The critical points of the potential function are indeed the equilibrium points of $\frac{dy}{dt}$, where the minima correspond to the stable equilibria, and the maxima correspond to the unstable equilibria. For the autonomous salinity difference in equation \ref{autonomous salinity difference dynamics}, it is simple to integrate against $y$ to find the potential function.

\[V(y) = m^2 ( \frac{\theta^2}{4}   y^4 - \frac{2\theta^3}{3}  y^3 + \frac{\theta^4}{2}y^2 ) + \frac{y^2}{2} - p^*y\]

\begin{figure}[H]
    \centering
    \includegraphics[width=0.5\linewidth]{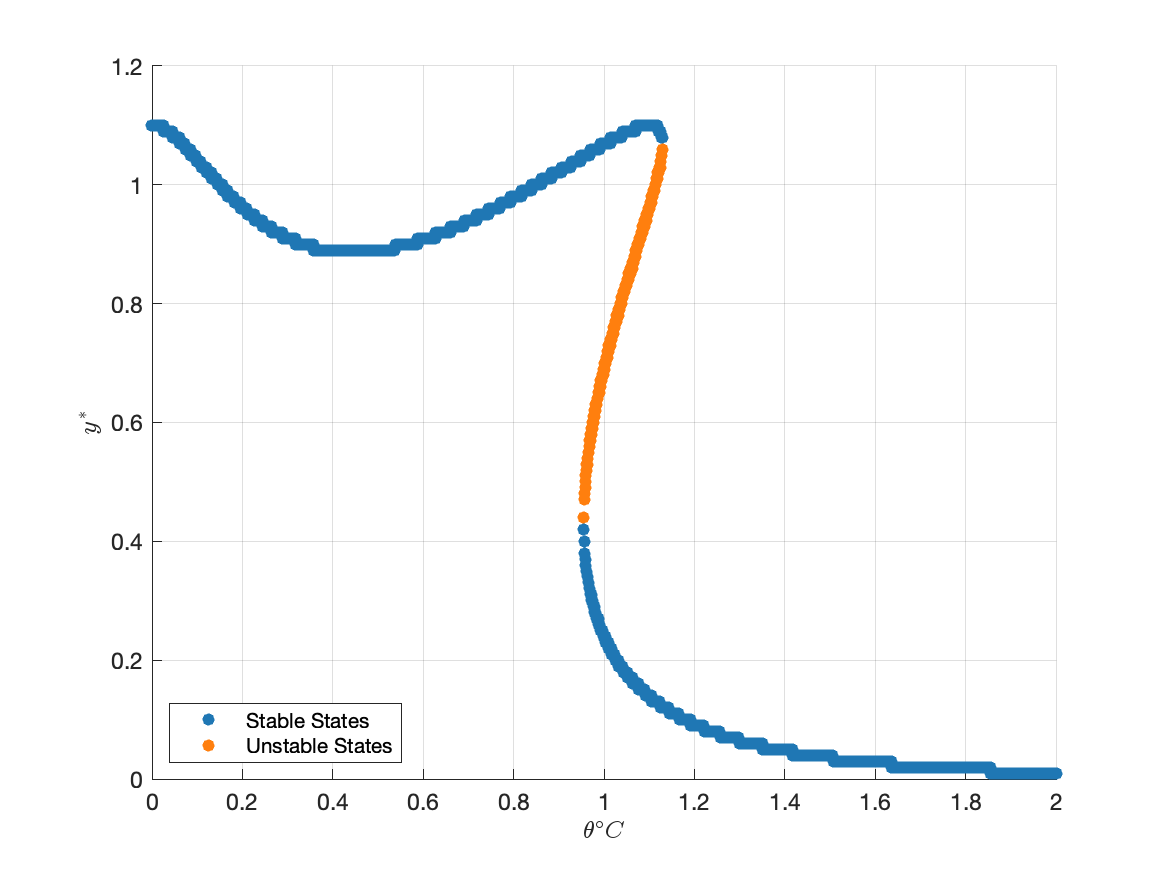}
    \caption{The depiction changes to stable and unstable states as a function of temperature gradient $\theta$, varied between 0 and 2. Blue points correspond to stable states while orange points correspond to unstable states. There are two saddle node bifurcations. The first occurs at $\theta=0.955$ and marks a structure transition from globally stable states around high salinity differences to multistability with two stable states and an unstable state. This structure continues until $\theta=1.13$, where there is a transition from multistability to global stability, centered around a low salinity difference. }
    \label{fig:bifurcation_temp}
\end{figure}

One can vary the temperature gradient $\theta$ to inspect changes in the potential landscape. We vary $\theta$ between 0 and 2 and computationally find a change in stability structure. In particular, a saddle node and a reverse saddle node bifurcation are found. The system is globally stable around a strong density difference for $\theta < 0.955$. At $\theta = 0.955$, a bifurcation occurs where the system becomes multistable with one equilibrium corresponding to a strong density difference, another stable equilibrium corresponding to a weaker density difference, and an unstable equilibrium.  At $\theta = 1.13$, the structure undergoes another bifurcation, where $\theta > 1.13$ corresponds to a globally stable tending towards a weak density difference. Figure \ref{fig:bifurcation_temp} summarizes the changes in structure of the system in a bifurcation diagram.

\begin{figure}[H]
    \centering
    \includegraphics[width=0.9\linewidth]{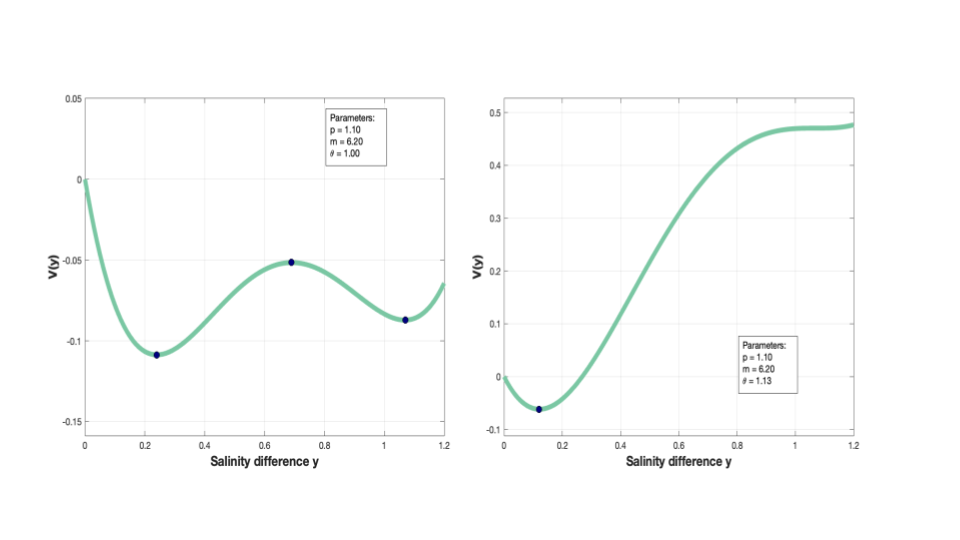}
    \caption{Depiction of potential landscapes as a function of the salinity difference $y$ under two parameter regimes. The panel on the left expresses the potential under freshwater forcing $p=1.1$, exchange rate 6.2, and temperature equilibrium $\theta=1$. Two local minima correspond to stable salinity difference states at 0.24 and 1.07, and one local maximum, corresponds to unstable state at 0.69.The panel on the right expresses the potential under increased temperature gradient $\theta=1.13$, with other parameters held the same as left panel. The increase in temperature gradient results in a structural change to only one local minima corresponding to a stable salinity difference state less than 0.67.}
    \label{fig:landscape_temp}
\end{figure}

Figure \ref{fig:landscape_temp} demonstrates the change in the potential landscapes by varying the temperature gradient regimes from the control system, where $\theta = 1$ to the bifurcation value $\theta=1.13$. We see that as $\theta$ increases, the equilibrium corresponding to the stronger density difference and the unstable equilibrium approach each other until they collide and annihilate each other at $\theta = 1.13$, which is the saddle node bifurcation seen in figure \ref{fig:bifurcation_temp}. The system tends towards a weak salinity difference beyond the bifurcation value.

\begin{figure}[H]
    \centering
    \includegraphics[width=0.5\linewidth]{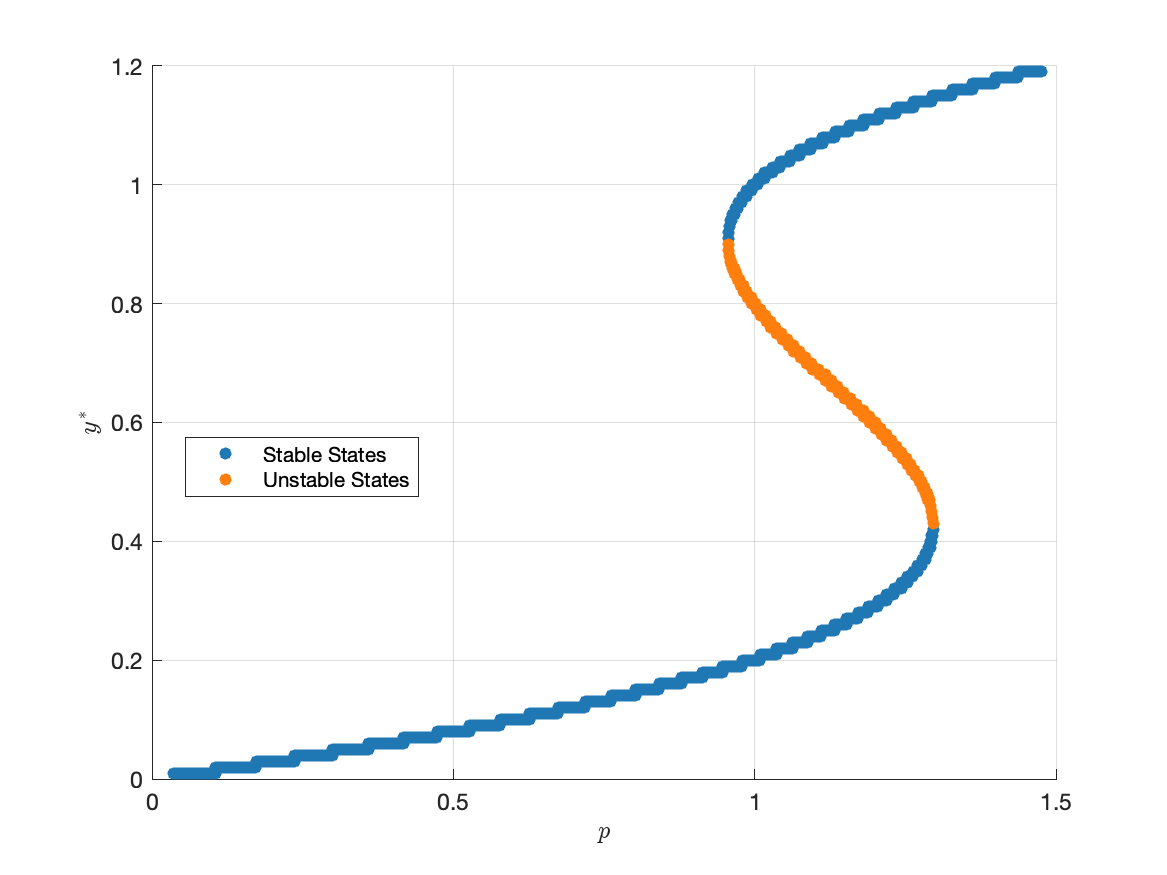}
    \caption{Depiction of stable and unstable states as a function of freshwater flux $p$, varied between 0 and 1.5. Blue points correspond to stable states while orange points correspond to unstable states. There are two saddle node bifurcations. The first occurs at $p = 0.95$ and marks a structure transition from globally stable states near low salinity differences to multistability, with two stable states and an unstable state. This structure continues until $\theta=1.07$, where there is a transition from multistability to global stability, centered around high salinity difference.}
    \label{fig:Cessi_Bifurcation_p}
\end{figure}

For comparison, one can obtain a similar diagram of Cessi's original potential function \ref{eqn: Cessi potential} under freshwater forcing. Varying the  parameter $p$ between 0 and 4, changes to the stability structure can be summarized by the bifurcation diagram in figure \ref{fig:Cessi_Bifurcation_p}. This figure verifies the critical value that Cessi identified in the freshwater influx, $p = 1.3$, which causes the system to transition from multistability to global stability, with the equilibrium corresponding to a stronger salinity difference than any salinity difference state in the multistable regime. The diagram also contributes the illumination of a reverse saddle node bifurcation: as the freshwater forcing is reduced, the system changes from multistable, within the range $0.95 < p < 1.3$, to globally stable, with flows tending towards a state of corresponding to a weakened salinity difference.

\subsection*{Discussion}

Motivated by observed increases in global mean temperature and by the observed polar amplification, we took a new look at Cessi's 1994 paper as a conceptual model that might provide insight into whether increases in the temperature gradient may cause a tipping in ocean circulation.  Starting with a variation of Stommel's two-box model, where the independent variables are the temperature and salinity gradients between the boxes, Cessi reduced the system to one variable by assuming that the temperature gradient relaxes to its equilibrium faster than the salinity gradient.  That assumption allowed her to hold the temperature gradient at its equilibrium value while letting the salinity gradient vary.  She thus reduced the system to single independent variable: the deviation of salinity from its equilibrium value.  Treating the influx of salinity into the polar box as a parameter, Cessi explored the tipping behavior of the system in response to a pulse of the parameter.

In this paper we examined the effect on Cessi's model of holding the pulse of freshwater constant while treating the equilibrium temperature gradient as the parameter of interest.  We found that the behavior is similar in quality to that in Cessi's paper: small changes in the parameter can induce large changes in the dynamic behavior of the system. The tendency toward a single stable state  corresponding to a weak salinity difference as a response to temperature gradient forcing differed from the tendency toward a  strong salinity difference as a response to freshwater forcing. 

\section*{Acknowledgments}
I am grateful for Dick McGehee, whose instruction in the Math 5490 course at University of Minnesota inspired the close inspection of this topic. Thank you to UMN Math Climate Seminar and the Mathematics of Climate Research Network for helpful discussions and mentorship. I would like to thank the members of my oral examination committee — Kate Meyer, Rick Moeckel, and Arnd Scheel — for their valuable feedback, thoughtful questions, and support throughout the examination process. ChatGPT-4o was used to correct spelling and grammar, and to improve phrasing for clarity. The writing phase of this project was funded by the Department of Education, GAANN fellowship.

\newpage
\bibliographystyle{siamplain}

\input{references.bbl}
\end{document}

%% file: ex_shared.tex
\usepackage{lipsum}
\usepackage{amsfonts,amsmath, amssymb}
\usepackage{graphicx}
\usepackage{epstopdf}
\usepackage{algorithmic}
\ifpdf
  \DeclareGraphicsExtensions{.eps,.pdf,.png,.jpg}
\else
  \DeclareGraphicsExtensions{.eps}
\fi

\newsiamremark{remark}{Remark}
\newsiamremark{hypothesis}{Hypothesis}
\crefname{hypothesis}{Hypothesis}{Hypotheses}
\newsiamthm{claim}{Claim}
\newsiamremark{fact}{Fact}
\crefname{fact}{Fact}{Facts}

\headers{Temperature induced tipping in a two box ocean circulation model}{J. Noory}

\title{Temperature induced tipping in a two box ocean circulation model\thanks{Submitted to the editors DATE.
\funding{This work was funded in part by US Department of Education GAANN Fellowship}}}

\author{Jasmine Noory\thanks{University of Minnesota 
  (\email{noory003@umn.edu}\url{}).}}

\usepackage{amsopn}
